\documentclass[twocolumn,prb,showpacs,preprintnumbers,amsmath,amssymb]{revtex4}

\usepackage{graphicx}
\usepackage{dcolumn}
\usepackage{bm}
\usepackage{epsfig}

\begin{document}


\title{Comparison of shear and dielectrics in highly viscous liquids}

\author{U. Buchenau}
\email{buchenau-juelich@t-online.de}
\affiliation{%
Institut f\"ur Festk\"orperforschung, Forschungszentrum J\"ulich\\
Postfach 1913, D--52425 J\"ulich, Federal Republic of Germany}%

\date{June 6, 2010}

\begin{abstract}
A relation between shear and dielectric spectra is derived for highly viscous liquids with a small rotational contribution $\Delta\epsilon$ to the dielectric constant. It is valid if the shear fluctuations and the electric dipole fluctuations have the same spectrum. The comparison to literature data, taken under carefully controlled conditions to ensure samples from the same charge and the same temperature control in both measurements, indicates that the relation may be fulfilled or not depending on the substance. The connection to recent work on strong correlations is discussed.
\end{abstract}


\maketitle

A promising approach for the study of the flow process in highly viscous liquids is its comparison \cite{donth,dgeba,carpentier,niss} in different techniques. One usually finds the dielectric absorption peak close to the heat capacity one \cite{donth,dgeba,carpentier}, but the shear modulus peak about half a decade higher in frequency. In a broad distribution of relaxation times, a modulus peak always appears at a higher frequency than a susceptibility (compliance) peak. The question is whether this is the reason here.

The dielectric susceptibility $\epsilon(\omega)$ is difficult to invert because it has two contributions, the electronic polarizability and the molecular dipole orientation. It is better to invert the shear modulus $G(\omega)$.

Let us start \cite{ferry} with the complex shear compliance
\begin{equation}\label{Jom}
J(\omega)=\frac{1}{G}+\int_{-\infty}^\infty \frac{L(\tau)}{1+i\omega\tau}d\ \ln\tau-\frac{i}{\omega\eta}.
\end{equation}

Here $G$ is the infinite frequency shear modulus and $\eta$ is the viscosity. A third material constant hidden in this equation is the recoverable compliance $J_e^0$, the elastic compliance plus the integral over the retardation processes
\begin{equation}
	J_e^0=\frac{1}{G}+\int_{-\infty}^\infty L(\tau)d\ \ln\tau.
\end{equation}

Eq. (\ref{Jom}) makes a separation of two contributions to the compliance, the retardation spectrum and the viscosity. From our gradually growing understanding of the highly viscous liquid \cite{stillinger,heuer}, we know that both parts must come from thermally activated transitions between inherent states, stable structures corresponding to minima of the potential energy. But the question is whether the retardation spectrum, which is only part of the response, can be used to calculate the normalized shear susceptibility from the fluctuation-dissipation theorem.

In order to argue that this is indeed the case, consider an inherent state $i$ of a small volume $V$ in the liquid. $V$ should be large enough to be characterized by the continuum viscoelasticity, but small compared to the sample into which it is embedded. The state $i$ has a well-defined electric dipole moment $\vec{p}_i$ and a well-defined shear tensor. For simplicity, we consider only one of the five possible shear directions. Let $e_i$ be the shear strain of state $i$ in this direction in the absence of external forces. The contribution $\delta_i$ of the state to the shear fluctuations of the volume $V$ in the given direction depends on the time-dependent actual zero point $e_0(t)$ of the strain of the surroundings, which can take all possible values in a liquid. The actual strain of the volume $V$ in the compromise of internal and external stresses is close to $(e_i-e_0(t))/2$, depending on the Poisson ratio \cite{eshelby}. For our argument, one can take it to be exactly half of the difference. Then, the free energy $E_i$ of the state $i$ has a stress contribution $GV(e_i-e_0)^2/4$ and 
\begin{equation}\label{di}
	\delta_i=f_i(e_i-e_0(t))^2/4
\end{equation}
where the occupation probability $f_i$ is given by
\begin{equation}\label{fi}
	f_i=\frac{1}{Z}\exp(-E_i/k_BT),
\end{equation}
$Z$ being the partition function of the volume $V$ for the given $e_0(t)$.

\begin{figure}[b]
\hspace{-0cm} \vspace{0cm} \epsfig{file=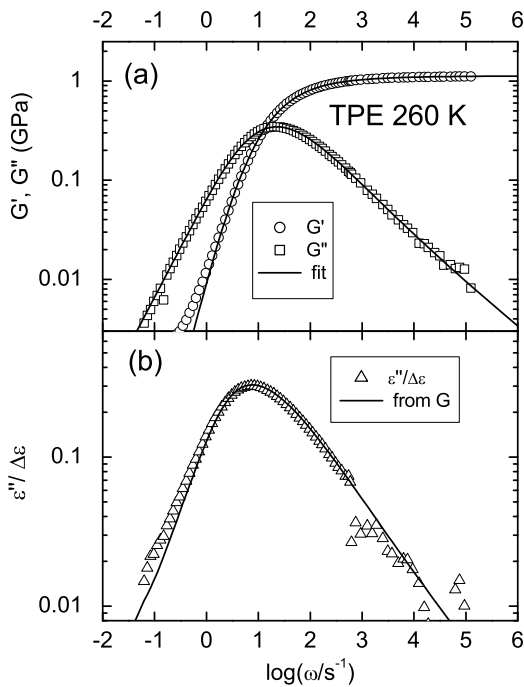,width=7cm,angle=0} \vspace{0cm}\caption{Validity of the relation in triphenylethylene (TPE) (a) Fit of dynamical shear data \cite{niss} at 260 K (b) Comparison of the measured dielectric damping at the same temperature to the one calculated from the shear fit.}
\end{figure}

\begin{figure}[b]
\hspace{-0cm} \vspace{0cm} \epsfig{file=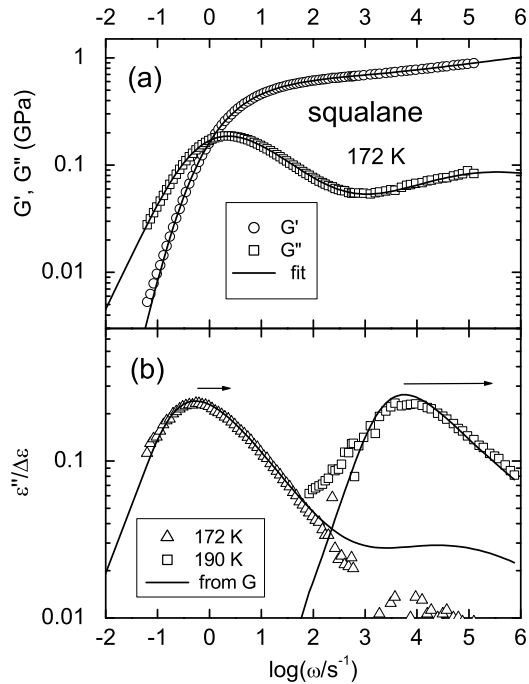,width=7cm,angle=0} \vspace{0cm}\caption{Validity of the relation in squalane (a) Fit of dynamical shear data \cite{niss} at 172 K (b) Comparison of the measured dielectric damping at 172 K and 190 K to the one calculated from the shear fit at the two temperatures, respectively. The arrows show the positions of the peak in $G''(\omega)$.}
\end{figure}

\begin{figure}[b]
\hspace{-0cm} \vspace{0cm} \epsfig{file=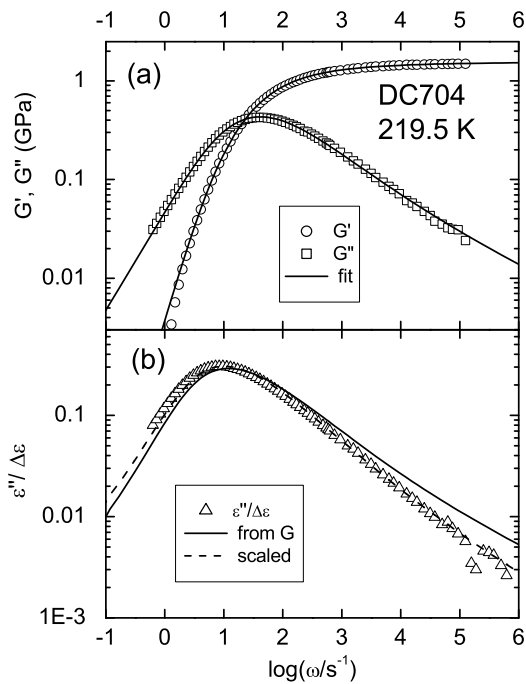,width=7cm,angle=0} \vspace{0cm}\caption{Peak shift in DC704 (a) Fit of dynamical shear data \cite{niss} at 219.5 K (b) Comparison of the measured dielectric damping at the same temperature to the one calculated from the shear fit (continuous line). The dashed line is obtained assuming a decrease of the dielectrics-shear coupling ratio by 6 \% per frequency decade.}
\end{figure}

\begin{figure}[b]
\hspace{-0cm} \vspace{0cm} \epsfig{file=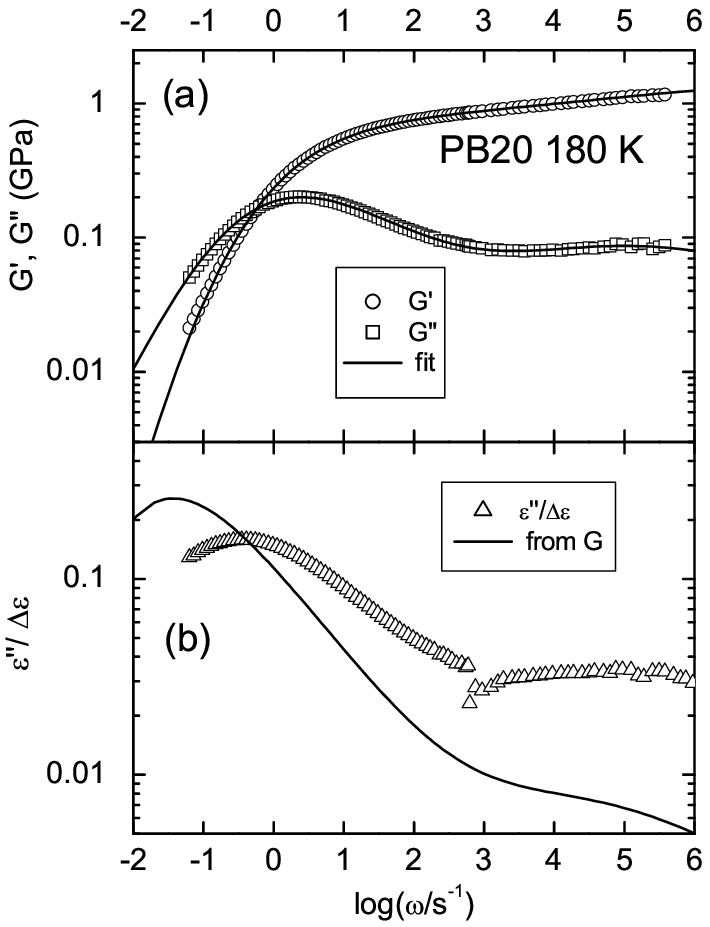,width=7cm,angle=0} \vspace{0cm}\caption{Strong differences in polybutadiene (a) Fit of dynamical shear data \cite{niss} at 180 K (b) Comparison of the measured dielectric damping at the same temperature to the one calculated from the shear fit.}
\end{figure}

In the case of a stationary flow $\dot{e}_0=const$, one has
\begin{equation}\label{deisdt}
	\frac{\partial E_i}{\partial t}=\frac{1}{2}GV(e_i-e_0(t))\dot{e}_0,
\end{equation}
from which one can calculate the contribution $\dot{e}_{0i}$ of state $i$ to the flow
\begin{equation}\label{de0i}
	\dot{e}_{0i}=\frac{GV}{2k_BT}\frac{1}{Z}f_i(e_i-e_0(t))^2\dot{e}_0.
\end{equation}

The contribution of the state to the viscosity is proportional to its contribution to the shear fluctuations, with a proportionality constant which is the same for all states. This intimate relation between viscosity and fluctuations shows that the viscous flow is not a separate process.

Thus the normalized shear susceptibility is defined by 
\begin{equation}\label{chis}
\chi_s(\omega)=\frac{G}{GJ_e^0-1}\ \int_{-\infty}^\infty \frac{L(\tau)}{1+i\omega\tau}d\ \ln\tau.
\end{equation}

With $J(\omega)=1/G(\omega)$ and eq. (\ref{Jom}), one can translate this to
\begin{equation}\label{chisa}
	\chi_s=\frac{G}{GJ_e^0-1}\left(\frac{1}{G(\omega)}-\frac{1}{G}+\frac{i}{\omega\eta}\right).
\end{equation}

$\chi_s$ is normalized with respect to the compliance step; the function begins with the value 1 at low frequency and ends with 0 at high frequency. To see this, note the limiting values of $G(\omega)$
\begin{equation}
	\lim_{\omega\rightarrow \infty}G'=G
\end{equation}
and
\begin{equation}
	\lim_{\omega\rightarrow 0}G'=J_e^0\eta^2\omega^2
\end{equation}
and
\begin{equation}
	\lim_{\omega\rightarrow 0}G''=\eta\omega
\end{equation}
(the last two are equs. (34) and (35) of chapter 3 of Ferry's book \cite{ferry}).

As long as the rotational contribution $\Delta\epsilon=\epsilon_{low}-\epsilon_{high}$ (where $\epsilon_{low}$ is the low frequency limit and $\epsilon_{high}$ is the high frequency limit of the dielectric constant) is small compared to 1, one can calculate the normalized complex dielectric susceptibility function $\chi_\epsilon$ from the measured complex $\epsilon(\omega)$ via
\begin{equation}\label{simple}
	\chi_\epsilon=\frac{\epsilon-\epsilon_{high}}{\Delta\epsilon}.
\end{equation}

If the flow relaxation processes affect shear strain and electric dipole fluctuations in the same way,
\begin{equation}\label{fin}
	\frac{\epsilon''}{\Delta\epsilon}=\frac{G}{GJ_e^0-1}\left(\frac{G''}{G'^2+G''^2}-\frac{i}{\omega\eta}\right).
\end{equation}

This equation will be used in the comparison to experiment \cite{niss}.

The comparison is done for four samples with a weak $\Delta\epsilon$: TPE (triphenylethylene), squalane, 1,4-polybutadiene and DC704, a silicon oil used in diffusion pumps \cite{niss}.

One way to compare is to fit $G(\omega)$, then calculate the expected normalized dielectric susceptibility from this fit and compare to the dielectric measurement. The fit is better than the data themselves, because small deviations in $G(\omega)$ at low $\omega$ tend to explode in the calculated retardation spectrum (see, for example, the increasing error bars in Figs. 3 and 4 of Schr\"oter and Donth's paper \cite{donth}). The fit also supplies the three parameters $G$, $\eta$ and $J_e^0$ needed for eq. (\ref{fin}). $\Delta\epsilon$ is the only parameter which has to be adapted to the dielectric data. 

The procedure does not remove the accuracy problem at low frequency completely; the position of the resulting dielectric $\alpha$-peak remains rather sensitive to the fit parameters for $G(\omega)$. The physical reason for this is the viscosity, which overshadows the low frequency relaxations in a shear measurement (but not in a dielectric measurement). Here, $G(\omega)$ was modeled in terms of the asymmetry model \cite{asymm}, introducing as an extra parameter a width $w$ of the cutoff at the critical barrier $V_c$ in order to get the best possible fit.

In TPE, the first example, the relation is fulfilled within experimental error. This is seen in Fig. 1. Fig. 1 (a) shows the fit to $G(\omega)$, Fig. 1 (b) the comparison of the dielectric measurement to the prediction. Other temperatures yield the same result.

The same good agreement is found in squalane, our second example. 
Fig. 2 (a) shows again the fit to $G(\omega)$ at 172 K, Fig. 2 (b) the comparison between predicted and measured dielectric damping at two temperatures, 172 and 190 K. This case is particularly impressive, because the shift between shear and dielectric peak increases from half a decade at the glass transition to one and a half at higher temperatures. Nevertheless, one finds good agreement at all temperatures, though the secondary peak is clearly weaker in dielectrics than in shear.

The third case, DC704 in Fig. 3, does not show the perfect agreement of the two previous cases. The calculated dielectric $\alpha$-peak lies a bit to the right of the measured one. However, one also finds a slightly higher negative slope in the $\epsilon''$ data, indicating a decrease of the ratio of the coupling constants of the elementary relaxations to dielectrics and shear with increasing frequency. If one corrects with a decrease of 6 \% per frequency decade, one recovers a good fit. Again, the result does not depend on temperature.

The last case, 1,4-polybutadiene (more precisely, a mixture of 80 \% 1,4-polybutadiene and 20 \% 1,2-polybutadiene) in Fig. 4 shows strong differences, with the calculated dielectric peak an order of magnitude slower than the observed one. Here, the deviations are much too pronounced to be repaired by the simple scaling which worked for DC704. However, this complete breakdown is in fact understandable. 1,4-polybutadiene consists of two kinds of monomers, {\it cis} and {\it trans}, with the {\it cis}-monomers responsible for the dielectric signal. A recent thorough simulation study \cite{narros} of 1,4-polybutadiene finds that their response is decidedly faster than the one of the {\it trans} monomers.

Our findings are not completely independent on our choice of a model \cite{asymm} for $G(\omega)$; taking other models, one can even fit $G(\omega)$ with a diverging recoverable compliance \cite{niss}. However, the results can be checked independently by an alternative procedure. One can fit the dielectric data first with a Cole-Davidson function (adding a Cole-Cole function for the secondary peak if necessary). With this fit, one calculates a normalized dielectric susceptibility from eq. (\ref{simple}). Then, one can check whether this susceptibility describes the $G(\omega)$ data correctly according to eq. (\ref{chisa}), taking not only $G$, $\eta$ and $J_e^0$ as free parameters, but also the time constant of the Cole-Davidson function. If this agrees with the one in the dielectric fit within experimental error (about a tenth of a decade) and if the fit is satisfactory, one again concludes that the shear and dielectric fluctuation spectra agree. In our four cases, this procedure corroborated the results from the $G(\omega)$-fit, thus supporting the choice of a shear model with a finite recoverable compliance.

Note that the identity of dielectric and shear retardation spectra found in two of our four cases has a different reason from the one of energy and density fluctuation spectra in strongly correlated liquids \cite{corr1,corr3} with a Prigogine-Defay-ratio close to one. In that case, the identity of the spectra is due to the strong correlation of the fluctuations. In our case, dielectric and shear fluctuations are not correlated at all, because a vector cannot be correlated with a tensor for symmetry reasons. Equal spectra do not necessarily imply strong correlation, though strong correlation implies equal spectra. In fact, the consideration on inherent states at the beginning of this paper shows clearly that the contributions of a given state to the shear strain and electric polarization fluctuations must be completely uncorrelated, because the first depends crucially on the external shear $e_0(t)$ and the second does not.

The equality of the spectra must rather result from the decay mechanism. If the local strain $e_i$ disappears completely in a transition between two inherent states, the same must be true for the local dipole moment $p_i$. In particular, the full decay of the shear strain should be accompanied by a full decay of the dielectric polarization (counterexamples are the normal modes in polymers or the monoalcohols \cite{bo}).

Summarizing the experimental and numerical evidence, one has two substances where eq. (\ref{fin}) is valid at all measured temperatures within experimental accuracy and two other substances where one has good reasons to attribute the deviations to the frequency dependence of the coupling ratio. While a final judgment requires the investigation of more cases, the results strongly favor the conclusion that shear fluctuation spectrum and dielectric fluctuation spectrum agree in simple substances and disagree if there is a complication like the one in 1,4-polybutadiene.

The finding provides for the first time a solid basis for the detailed comparison of the two most important techniques in the study of highly viscous liquids. If the spectra agree, the two techniques are complementary, the dielectric data providing the exact shape of the spectrum, in particular at the low frequency end which is not well seen in the shear measurement, and the shear data supplying the three material constants shear modulus, viscosity and recoverable compliance which together with the spectrum describe the flow. If the spectra do not agree, one can look for the microscopic reason.

The paper profited a lot from intense discussions with the Roskilde group, in particular Niels Boie Olsen, Bo Jakobsen, Kristine Niss, Tage Christensen, Albena Nielsen and Tina Hecksher.

\end{document}